\documentclass[showpacs,preprintnumbers,prd,showkeys,aps,twocolumn]{revtex4}

\usepackage{eurosym}
\usepackage{amssymb}
\usepackage{amsfonts}
\usepackage{amsmath}
\usepackage{graphicx}
\usepackage{changes}
\usepackage{calrsfs}
\usepackage[colorlinks=true,
            linkcolor=red,
            urlcolor=blue,
            citecolor=blue]{hyperref}

\begin{document}

\title{Light Deflection by a Quantum Improved Kerr Black Hole Pierced by a Cosmic String
 }
\author{Kimet Jusufi}
\email{kimet.jusufi@unite.edu.mk}
\affiliation{Physics Department, State University of Tetovo, Ilinden Street nn, 1200, Tetovo, 
Macedonia}
\affiliation{Institute of Physics, Faculty of Natural Sciences and Mathematics, Ss. Cyril and Methodius University, Arhimedova 3, 1000 Skopje, North Macedonia}

\author{Ali \"{O}vg\"{u}n}
\email{ali.ovgun@pucv.cl}

\affiliation{Instituto de F\'{\i}sica, Pontificia Universidad Cat\'olica de
Valpara\'{\i}so, Casilla 4950, Valpara\'{\i}so, Chile}

\affiliation{Physics Department, Faculty of Arts and Sciences,
Eastern Mediterranean University, Famagusta, 99628, North Cyprus, via Mersin 10, Turkey.}

\date{\today }

\begin{abstract}
In this work, we calculate the quantum correction effects on the deflection of light in the spacetime geometry of a quantum improved Kerr black hole pierced by an infinitely long cosmic string. More precisely, we calculate the deflection angle by applying the Gauss--Bonnet theorem (GBT) to the osculating optical geometries related to the quantum improved rotating black hole in the weak limit approximation. We find that the deflection angle of light is affected by the quantum effects as well as the global topology due to the presence of the cosmic string. Besides, we have managed to find the same expression for the deflection angle in leading order terms using the geodesic equations. 
\end{abstract}

\maketitle

\section{Introduction}

Einstein's theory of relativity predicts that light rays gets deflected due to the curvature of spacetime known as the gravitational lensing effect. The deflection of light has been extensively studied previously in the literature in various astrophysical systems in the context of the weak as well as strong gravitational lensing limit \cite{weak1,weak2,strong1,strong2,strong3,virbhadra1,virbhadra2,virbhadra3,sereno1,sereno2,sereno3}.

A very important contribution has been recently made by Gibbons and Werner who argued about the importance of global topology on the deflection of light using the optical geometry and the famous GBT \cite{gibbons1}.  Furthermore, they computed the deflection angle from the Schwarzschild black hole  by considering a domain outside of the light ray - different from the standard method where the lensing effect is strongly related with the mass of the body which is enclosed within a given region on space. 
More recently, Werner has extended this method and computed the deflection angle by a Kerr black hole \cite{werner}. Amazingly, this method was shown to be very suitable in calculating the deflection angle in spacetimes with topological defects, including cosmic strings and global monopoles \cite{kimet1,kimet2,kimet3}. Quantum effects on the deflection of light by a quantum improved Schwarzschild black holes and gravitational effects due to a cosmic string in Schwarzschild spacetime \cite{kimet4,ao}.

In this paper, we will use the Gibbons-Werner (GW) method to obtain deflection angle in a quantum improved Kerr black hole geometry with a cosmic string. Topological defects are associated with a number of quantum and gravitational effects, including early structure formation from cosmic string loops which have been studied by Benjamin et al. in Ref. \cite{benjamin}, light deflection by a cosmic string and global monopoles in Ref. \cite{kimet2,kimet3}. Recently, Hackmann et al. \cite{kerr1}, investigated the deflection of light by the Kerr black hole pierced by a cosmic string. In this paper, using similar arguments, we shall introduce a quantum improved Kerr black hole metric pierced by a static and infinitely long cosmic string lying along the $z$ --axis to calculate the deflection angle in the weak approximation limit.  Various modifications on the deflection of light and also modifications in the context of quantum gravity effects \cite{bohr}, non-linear electrodynamics \cite{novello}, have been studied in the context of some alternative gravity theories \cite{sumanta1,sumanta2}, in particular deflection in the strong limit by Eddington-inspired Born-Infeld black holes \cite{Born-Infeld1,Born-Infeld2}.

Classically, a black hole has a horizon and also a singularity, however people have tried to remove singularities from black holes using various methods such as nonlinear electrodynamics, modified gravities or, using the effects of the quantum gravity. In this paper, we wish to extend the Kerr solution by taking into account the quantum effects into the deflection of light in the spacetime of a quantum improved Kerr black hole recently found by Torres \cite{torres}. The main idea to recover this metric is a running Newton's constant $G=G(k)$ \cite{qcsh0}. 

Further, let us now briefly review the GBT which connects the topologically surfaces. First, using Euler characteristic $\chi$ and a Riemannian metric $g$, one can choose the subset oriented surface domain as $(D,\chi,g)$ to find the Gaussian curvature $K$. Then the Gauss-Bonnet theorem is defined as follow \begin{equation}
\int\int_D K \rm d S + \int_{\partial D} \kappa \rm d t +\sum_i \alpha_i = 2\pi \chi(D).
\label{gb}
\end{equation} where $\kappa$ is the geodesic curvature for  $\partial D:\{t\}\rightarrow D$ and $\alpha_i$ is the exterior angle with $i^{th}$ vertex. Following this approach, global symmetric lenses are considered to be Riemannian metric manifolds, which are geodesic spatial light rays. In optical geometry, we calculate the Gaussian optical curvature $K$ to find the asymptotic bending angle which can be calculated as follows:
\begin{equation}
\hat{\alpha}=-\int \int_{D_\infty} K  \mathrm{d}S.
\end{equation}

Note that this equation is an exact result for the deflection angle. In this equation, we integrate over an infinite region of the surface $D_\infty$ which is bounded by the light ray. By assumtion, Aone can use the above relation only for asymptotically Euclidian optical metrics. Therefore it will be interesting to see the form of the deflection angle in the case of non-asymptotically Euclidian metrics.  One such interesting metric is to consider the Kerr black hole pierced by a cosmic string. 
We calculated the deviation aberration by taking a straight line as the zeroth approximation of the light ray. With this method deflection angle $\hat{\alpha}$ can be easily found but only in leading order terms. The main difficulty here is that the optical geometry should transform into a special type of Finsler manifold, known as the Randers manifold. Then we shall use the Naz{\i}m's method to construct a Riemannian manifold osculating the Randers manifold.

This paper is organized as follows. In Section II, we first recall the quantum improved rotating black hole solution, after that we introduce a cosmic string in this spacetime with the corresponding quantum improved Kerr-Randers-Cosmic string optical metric. In Section III,  we calculate the quantum improved Gaussian optical curvature which is used to calculate the deflection angle by applying GBT to the quantum improved optical geometry. In Section IV, we study the geodesic equations in the spacetime of quantum improved Kerr black hole with a cosmic string. In particular we calculate the deflection angle in leading order. In Section V, we elaborate on our results.

\section{Quantum improved Kerr-Randers-Cosmic string optical metric}

Recently a non-singular solution, also known as a quantum improved rotating black hole  was found by Torres \cite{torres,qcsh0}. The basic idea is to use the running Newton's constant $G=G(k)$ with a position-dependent scale parameter i.e. $k=k(r)$ in the asymptotic safety approach. The quantum improved rotating black hole metric was found to be
\begin{eqnarray}\label{1}\notag
\mathrm{d}s^{2}&=& - \frac{\Delta_{\tilde{\omega}}}{\Sigma}\left(\mathrm{d}t-a \sin^2 \theta \mathrm{d}\varphi \right)^2+\frac{\Sigma}{\Delta_{\tilde{\omega}}}\mathrm{d}r^2+\Sigma \mathrm{d}\theta^2\\
&+& \frac{\sin^2 \theta}{\Sigma}\left(a \mathrm{d}t-(r^2+a^2)\mathrm{d}\varphi \right)^2, \label{m1}
\end{eqnarray}
where 
\begin{equation}
\Delta_{\tilde{\omega}}=r^2+a^2-2G(r) M r,
\end{equation}
\begin{equation}
\Sigma=r^2+a^2 \cos^2 \theta,
\end{equation}
\begin{equation}
G(r)=\frac{G_{0}r^{3}}{r^{3}+\tilde{\omega} G_{0}\left(r+\gamma G_{0}M\right)}.
\end{equation}

Note that $k_{obs}$ gives a typical observational scale such that $G_0=G(k_{obs})$. The quantum effects are encoded by the parameter
\begin{equation}
\tilde{\omega}=\frac{167 \,\hbar}{30 \pi}.
\end{equation}
 that when the quantum effects are zero ($\tilde{\omega}=0$, the metric of Eq. (\ref{m1}) reduces to Kerr spacetime. Moreover, for the case of $a=0$, it becomes Schwarzschild spacetime.

Let us now introduce a cosmic string piercing of this Kerr black hole solution by using the following coordinate transformation \cite{kerr0,kerr1,kerr2}
\begin{equation}
\mathrm{d} \varphi \to \beta \, \mathrm{d} \varphi,
\end{equation}
where the cosmic string parameter is given as $\beta =1-4 \mu$, where $\mu$ is known as the energy density of the cosmic string. In this case, the spacetime (\ref{m1}) can be written as 
\begin{eqnarray}\notag
\mathrm{d}s^{2}&=& - \frac{\Delta_{\tilde{\omega}}}{\Sigma}\left(\mathrm{d}t-a \beta \sin^2 \theta \mathrm{d}\varphi \right)^2+\frac{\Sigma}{\Delta_{\tilde{\omega}}}\mathrm{d}r^2+\Sigma \mathrm{d}\theta^2\\
&+& \frac{\sin^2 \theta}{\Sigma}\left(a \mathrm{d}t-(r^2+a^2)\beta \mathrm{d}\varphi \right)^2. \label{metric}
\end{eqnarray}

Note that cosmic string parameter belongs to the interval $0<\beta <1$, while the deficit angle is given as $\delta=2\ \pi (1- \beta)$. In what follows we will use  the metric (\ref{metric}) to find out the  deflection angle of light. To do so, let us find the corresponding Finsler metric for our improved Kerr metric with a cosmic string. A Finsler metric $F$ with manifold $\mathcal{M}$ and $x\in \mathcal{M},\ X\in T_x M$ is characterised by the Hessian given as
\begin{equation} 
g_{ij}(x,X)=\frac{1}{2}\frac{\partial^{2}F^{2}(x,X)}{\partial X^{i}\partial X^{j}}.\label{10-3}
\end{equation}

Moreover the Randers metric is given as
\begin{equation}
F(x, X)=\sqrt{a_{ij}(x)X^{i}X^{j}}+b_{i}(x)X^{i},\label{11-3}
\end{equation}
with the following condition $a^{ij}b_{i}b_{j}<1$, with $a_{ij}$ being a Riemannian metric and $b_{i}$ being a one-form. 

Then our quantum improved Kerr metric with a cosmic string can be easily written as the following stationary form metric \cite{gibbons2}
\begin{equation}\label{13-3}
\mathrm{d}s^2=V^2\left[-\left(\mathrm{d}t-b_i \mathrm{d}x^i \right)^2+a_{ij}\mathrm{d}x^i \mathrm{d}x^j\right],
\end{equation}
where $V$ should be properly chosen.  If we consider null geodesics $\mathrm{d}s^2=0$, the last metric can be written in the form \eqref{11-3}, where
\begin{widetext}
\begin{eqnarray}
a_{ij}(x)\mathrm{d}x^i \mathrm{d}x^j&=&\frac{\Sigma^4}{\Delta_{\tilde{\omega}}-a^2 \sin^2 \theta}\left( \frac{\mathrm{d}r^2}{\Delta_{\tilde{\omega}}}+\mathrm{d}\theta^2+\frac{\Delta_{\tilde{\omega}} \sin^2\theta \beta^2 } {\Delta_{\tilde{\omega}}-a^2 \sin^2\theta }\mathrm{d}\varphi^2 \right),\\
b_{i}(x)\mathrm{d}x^i &=& -\frac{2 a M G(r) r \beta \sin^2 \theta}{\Delta_{\tilde{\omega}}-a^2 \sin^2 \theta}\mathrm{d}\varphi.
\end{eqnarray}

Hence, in the equatorial plane $\theta=\pi/2$, we find the corresponding quantum improved Kerr-Randers-String optical metric given by
\begin{equation}\label{16-3}
F\left(r,\varphi,\frac{\mathrm{d}r}{\mathrm{d}t},\frac{\mathrm{d}\varphi}{\mathrm{d}t}\right)=\sqrt{\frac{r^4 \beta^2 \Delta_{\tilde{\omega}} }{(\Delta_{\tilde{\omega}}-a^2)^2}\left(\frac{\mathrm{d}\varphi}{\mathrm{d}t}\right)^2+\frac{r^4 }{\Delta_{\tilde{\omega}}(\Delta_{\tilde{\omega}}-a^2)}\left(\frac{\mathrm{d}r}{\mathrm{d}t}\right)^2}-\frac{2M G(r)\beta ar}{\Delta_{\tilde{\omega}}-a^2}\frac{\mathrm{d}\varphi}{\mathrm{d}t}.
\end{equation}
\end{widetext}

One can immediately notice that $F$ can be used to describe the propagation of light after we first let $\mathrm{d}s^2=0$ which implies that $\mathrm{d}t=F(x,\mathrm{d}x)$. Fermat's principle in general relativity tells us that light rays $\gamma$ are choosen such that the following condition is satisfied
\begin{equation}
0=\delta\,\int\limits_{\gamma}\mathrm{d}t=\delta\,\int\limits_{\gamma_F}F(x, \dot{x})\mathrm{d}t.
\end{equation}

Where $\gamma_F$ is geodesic of our Kerr-Randers-Cosmic string optical metric $F$. The key idea here is to construct a Riemannian manifold $(\mathcal{M}\bar{g})$ which osculates the Randers manifold $ (\mathcal{M}, F) $ using the so-called Naz{\i}m's method \cite{nazim}. One way to do this, is to choose a vector field $\bar{X}$ tangent to the geodesic $\gamma_{F}$, such that $\bar{X}(\gamma_{F})=\dot{x}$, with the Hessian 
\begin{equation}
\bar{g}_{ij}(x)=g_{ij}(x,\bar{X}(x)).\label{17-3}
\end{equation}

Amazingly, one can check in Ref. \cite{werner}, that the geodesic $\gamma_{F}$ of the Randers manifold is also a geodesic $\gamma_{\bar{g}}$ of $(\mathcal{M},\bar{g})$, i.e.  $\gamma_{F}=\gamma_{\bar{g}}$. Thus, one can use the osculating Riemannian manifold $(\mathcal{M},\bar{g})$  to compute the deflection angle of light. Let us choose the undeflected light rays as $r(\varphi)=b/\sin\varphi $  with $b$ being the impact parameter which is approximated as the minimal distance to the cosmic string passing through the black hole. Near the light ray, we choose our vector field as follows
\begin{eqnarray} \nonumber\label{vec}
\bar{X}^{r}&=&-\cos\varphi+\mathcal{O}(M,a), \\
\bar{X}^{\varphi}&=&\frac{\sin^{2}\varphi}{b}+\mathcal{O}(M,a).
\end{eqnarray}

In the next section we shall preside to apply the GBT to the osculating optical geometry $(\mathcal{M},\bar{g})$ and compute the deflection angle. 

\bigskip
\section{Quantum improved Gaussian optical curvature and quantum improved deflection angle}

Let us choose a non-singular domain $(D_{R},\bar{g})$ over the osculating Riemannian manifold $(\mathcal{M},\bar{g})$ bounded by circular curve $C_ {R}$ and the geodesic $\gamma_{\bar{g}}$, such that $\partial D_{R}=\gamma_{\bar{g}}\cup C_ {R}$ (see Fig. 1). 

Then GBT can be stated as follows  (cf. \citep{werner})
\begin{equation}\label{19-4}
\iint\limits_{D_{R}}K\,\mathrm{d}S+\oint\limits_{\partial D_{R}}\kappa\,\mathrm{d}t+\sum_{i}\theta_{i}=2\pi\chi(D_{R}), 
\end{equation}
in which we note that $K$ is our quantum improved Gaussian curvature, $\kappa=|\nabla_{\dot{\gamma}}\dot{\gamma}|$ is the geodesic curvature, and $\theta_{i}$ being the $i^{th}$ exterior angles. In the limit $R\to \infty$, the sum of exterior jump angles at $S$ and $O$ gives $\theta_{O}+\theta_{S}\to \pi$. The Euler characteristic is $\chi(D_{R})=1$ since $D_ {R} $ is non-singular and simply connected. 

The GB theorem (\ref{19-4}) now can be written as,
\begin{equation} \label{gb2}
\iint\limits_{D_{R}}K\,\mathrm{d}S+\oint\limits_{\partial D_{R}}\kappa\,\mathrm{d}t=2\pi\chi(D_{R})-(\theta_{O}+\theta_{S})=\pi.
\end{equation}

\begin{figure}[h!]
\includegraphics[width=0.47\textwidth]{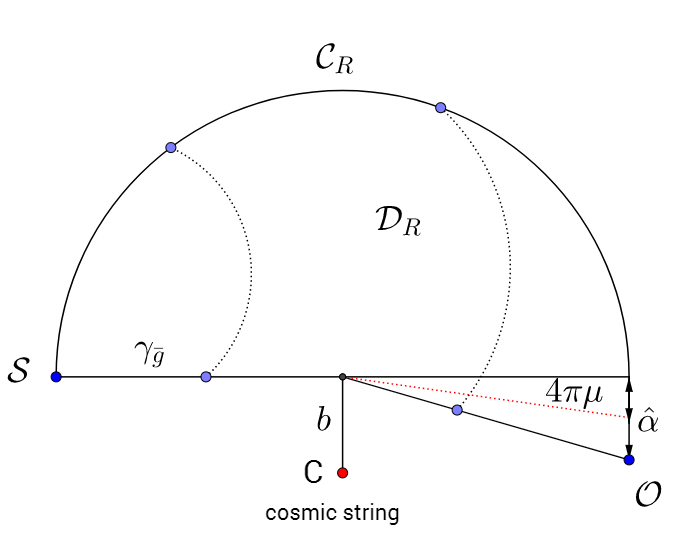} 
\caption{\small \textit {Deflection of light by a quantum improved Kerr black hole pierced by a cosmic string which is perpendicular to the equatorial plane $(r,\varphi)$. Due to the conical topology the deflection angle of light is $4 \pi \mu$, while the total deflection angle is $\hat{\alpha}$, the impact parameter $b$ is the minimal radial distance of the light ray from the cosmic string. Note that the vector field $\bar{X}(r,\varphi)$ is tangent to the geodesic. } }
\end{figure}

However we note that since $\gamma_{\bar{g}}$ is geodesic one is left with $\kappa(\gamma_{\bar{g}})=0$. Hence we must calculate $\kappa(C_{R})\mathrm{d}t$ where $\kappa(C_{R})=|\nabla_{\dot{C}_{R}}\dot{C}_{R}|$. Since $R$ is chosen such that  $C_{R}:= r(\varphi)=R=\text{const}$, only the radial component remains to be discussed
\begin{equation}
\left(\nabla_{\dot{C}_{R}}\dot{C}_{R}\right)^{r}=\bar{\Gamma}^{r}_{\varphi \varphi}\left(\dot{C}_{R}^{\varphi}\right)^{2}.
\end{equation}

Next, if we recall the unit speed condition given as $\bar{g}_{\varphi \varphi}\,\dot{C}_{R}^{\varphi}\dot{C}_{R}^{\varphi}=1$, and Christoffel symbol $\bar{\Gamma}^{r}_{\varphi \varphi}$, and choose  $r(\varphi)=R=\text{const}$, then for the geodesic curvature we find the following result $\kappa(C_{R})\to R^{-1}$. Note that this result is similar to the Kerr black hole without a cosmic string, however if one recall Eq. (\ref{16-3}) it follows that 
\begin{eqnarray}
\mathrm{d}t&=&\left[\sqrt{\frac{R^4 \beta^2 \Delta_{\tilde{\omega}} }{(\Delta_{\tilde{\omega}}-a^2)^2}}-\frac{2aM G(R)\beta  R}{\Delta_{\tilde{\omega}}-a^2}\right]\mathrm{d}\varphi.
\end{eqnarray}

This, of course, is a result of the global conical topology due to the presence of the cosmic string in our spacetime. As a result we have
\begin{eqnarray}\notag
& &\lim_{R \to \infty} \kappa(C_{R})\mathrm{d}t\\ \notag
&=&\lim_{R \to \infty}\left[\sqrt{\frac{R^2 \beta^2 \Delta_{\tilde{\omega}} }{(\Delta_{\tilde{\omega}}-a^2)^2}}-\frac{2aM G(R)\beta  }{\Delta_{\tilde{\omega}}-a^2}\right]\mathrm{d}\varphi \\
&\to & \beta d \varphi.
\end{eqnarray}

The last result suggests, due to the cosmic string our quantum improved optical geometry is different form the Kerr optical geometry in the sense that cannot be viewed as asymptotically Euclidean. In other words we have $\kappa(C_{R})\mathrm{d}t/\mathrm{d}\varphi=\beta\neq 1$, but reduces to  asymptotically Euclidean if and only if we let $\beta \to 1$. From the GB theorem (\ref{gb2}), one finds that
\begin{eqnarray}\notag
\iint\limits_{D_{R}}K\,\mathrm{d}S & + & \oint\limits_{C_{R}}\kappa\,\mathrm{d}t\overset{{R\to \infty}}{=}\iint\limits_{
D_{\infty}}K\,\mathrm{d}S \\
&+& \beta \int\limits_{0}^{\pi+ \hat{\alpha}}\mathrm{d}\varphi,
\end{eqnarray}
where we shall integrate over a  domain $D_\infty$ outside the light ray $\gamma_{\bar{g}}$. Morover $\hat{\alpha}$ is nothing else but the deflection angle of light to be calculated. After some algebraic manipulation the deflection angle is found to be
\begin{eqnarray}\label{alp}
\hat{\alpha} &\simeq & 4 \pi \mu - \frac{1}{1-4 \mu} \int\limits_{0}^{\pi}\int\limits_{\frac{b}{\sin \varphi}}^{\infty}K\,\sqrt{\det \bar{g}}\,\mathrm{d}r\,\mathrm{d}\varphi.
\end{eqnarray}

In what follows we shall compute the quantum improved Gaussian optical curvature, $K$. But first, we must find the corresponding metric components $\bar{g}$ of the osculating Riemannian geometry. Making use of the Hessian (\ref{10-3}), and Eq. (\ref{vec}), we find the following relations
\begin{widetext}
\begin{eqnarray}
\bar{g}_{rr}&=&\frac{-2 a \beta^3  Mr^3 \sin^6 \varphi G_0+\left(r^2 \beta^2 \sin^4 \varphi+\cos^2 \varphi \, b^2 \right)^{3/2}(4r G_0 M+r^2+G_0\tilde{\omega})}{\left(r^2 \beta^2 \sin^4 \varphi+\cos^2 \varphi \,b^2 \right)^{3/2} (G_0 \tilde{\omega}+r^2)},\\\notag
\bar{g}_{\varphi \varphi}&=&\frac{r^2 \beta^2 }{(G_0\tilde{\omega}+r^2)\left(r^2 \beta^2 \sin^4 \varphi+\cos^2 \varphi \,b^2 \right)^{5/2}}\\\notag
&\times & \Big[\Big((Mr G_0+\tilde{\omega}G_0+r^2)r^2 \beta^2 \sin^4\varphi  +\cos^2\varphi b^2  (MrG_0+r^2/2+\tilde{\omega}G_0/2)\Big)2 \left(r^2 \beta^2 \sin^4 \varphi+\cos^2 \varphi b^2  \right)^{3/2}\\\notag
&-&\Big(r \beta  \sin^2 \varphi (G_0\tilde{\omega}+r^2)\left(r^2 \beta^2 \sin^4 \varphi+\cos^2 \varphi b^2 \right)^{1/2}+aMG_0  (4 \beta^2 \sin^4 \varphi r^2+6 \cos^2 \varphi b^2  )\Big)\\ 
& \times & r \beta \sin^2 \varphi \left(r^2 \beta^2  \sin^4 \varphi+\cos^2 \varphi b^2 \right)\Big],\\
\bar{g}_{r\varphi}&=&\frac{2 a \beta MG_0 r \cos^3\varphi}{(G_0\tilde{\omega}+r^2)\left(\frac{r^2 \beta^2 \sin^4 \varphi+\cos^2 \varphi \,b^2 }{b^2}\right)^{3/2}},
\end{eqnarray}
with the determinant given as
\begin{equation}
\det \bar{g}= r^2 \beta^2 +\frac{6 M \beta^2 r^3 G_0}{G_0 \tilde{\omega}+r^2}-\frac{6 a \beta^3 M G_0 \sin^2\varphi r^3 }{\sqrt{r^2 \beta^2 \sin^4 \varphi+\cos^2 \varphi \,b^2 } \left(G_0 \tilde{\omega}+r^2\right)}+\mathcal{O}(a^2,M^2).
\end{equation}

The quantum improved Gaussian optical curvature is defined as
\begin{eqnarray}\label{25}
K=\frac{\bar{R}_{r\varphi r\varphi}}{\det \bar{g}}=\frac{1}{\sqrt{\det \bar{g}}}\left[\frac{\partial}{\partial \varphi}\left(\frac{\sqrt{\det \bar{g}}}{\bar{g}_{rr}}\,\bar{\Gamma}^{\varphi}_{rr}\right)-\frac{\partial}{\partial r}\left(\frac{\sqrt{\det \bar{g}}}{\bar{g}_{rr}}\,\bar{\Gamma}^{\varphi}_{r\varphi}\right)\right].
\end{eqnarray}

Using the above relations we find the following result
\begin{eqnarray}
K&=&-\frac{2 M G_0}{r^3 }+\frac{12 G_0^2 M \tilde{\omega}}{r^5\ }+\frac{6 a M G_0 f(r,\varphi,\tilde{\omega})}{r^9 },
\end{eqnarray}
where
\begin{eqnarray}\notag
f(r,\varphi,\tilde{\omega})&=& \frac{\sin^2 \varphi(r^2-3 G_0 \tilde{\omega})\beta }{\left(r^2 \beta^2 \sin^4 \varphi+\cos^2 \varphi \, b^2 \right)^{7/2}}\Big[r^8 \beta^6 (r^2-\frac{5 G_0 \tilde{\omega}}{3}) \sin^{12} \varphi-\frac{b^2 r^4 \beta^2 (G_0 \tilde{\omega}+r^2)^2 \sin^{10} \varphi}{2}\\\notag
&+&\frac{\cos^2\varphi \beta^2}{2} \left[(5\beta^2-9)r^4-142\tilde{\omega}G_0 r^2 (\beta^2+\frac{27}{22})- G_0^2 \tilde{\omega}^2 (\beta^2+9)\right] b^2 r^4 \sin^8 \varphi\\\notag
&+& 4 b^3 \beta^2 \cos^2\varphi (r^2+\frac{G_0 \tilde{\omega}}{2})(G_0\tilde{\omega}+r^2) r^3 \sin^7 \varphi+2 b^2 \cos^2\varphi  (b^2-\frac{5 \beta^2 \cos^2\varphi r^2}{2})(G_0\tilde{\omega}+r^2)^2 r^2 \sin^6 \varphi\\\notag
&+& 8 b^3 \cos^4\varphi \beta^2 (r^2+\frac{G_0 \tilde{\omega}}{2})(G_0\tilde{\omega}+r^2) r^3 \sin^5 \varphi-3 \beta^6 b^6 G_0 \tilde{\omega} \cos^6 \varphi (r^2-\frac{G_0 \tilde{\omega}}{3})\\\notag
&+&\cos^4\varphi \beta^2 \left[(\beta^2-\frac{11}{2})r^4+\frac{41}{3} \tilde G_0r^2{\omega}(\beta^2-\frac{33}{41})+2G_0^2 \tilde{\omega}^2(\beta^2-\frac{11}{4})\right]b^4 r^2 \sin^4 \varphi\\\notag
&+&5 b^4  \cos^6\varphi (G_0\tilde{\omega}+r^2)^2 r^2 \sin^2 \varphi- 2 b^5  \cos^6\varphi (G_0\tilde{\omega}+r^2) (3G_0\tilde{\omega}+r^2)r \sin \varphi \\
&-& b^5  \cos^4\varphi (G_0\tilde{\omega}+r^2) (3G_0\tilde{\omega}+r^2)r \sin^3 \varphi\Big].
\end{eqnarray}

Going back to Eq. (\ref{alp}) and substituting the quantum improved Gaussian optical curvature, the deflection angle can be calculated from the integral 
\begin{eqnarray}
\hat{\alpha} &\simeq & 4 \pi \mu-\frac{1}{1-4 \mu}\int\limits_{0}^{\pi}\int\limits_{\frac{b}{\sin \varphi}}^{\infty}\left(-\frac{2 M G_0}{r^3 }+\frac{12 G_0^2 M \tilde{\omega}}{r^5 }+\frac{6 a M G_0 f(r,\varphi,\tilde{\omega})}{r^9}\right)\,\sqrt{\det \bar{g}}\,\mathrm{d}r\,\mathrm{d}\varphi.
\end{eqnarray}

Integrating and using a Taylor series expansion near $\eta$ and $\hbar$ the deflecton angle results with
\begin{equation} \label{alp2}
\hat{\alpha}\simeq  4 \pi \mu +\frac{4 G_0 M}{b}+\frac{16 M G_0\mu}{b}-\frac{1336 G_0^2 M\hbar }{45 \pi b^3 }-\frac{5344 G_0^2 M \hbar \mu}{45 \pi b^3}\pm \frac{4 G_0 M a}{b^2}.
\end{equation}

We note that in the last equation we have used positive sign for retrograde and negative sign for prograde light rays. Furthermore we recover the Kerr black hole as a limiting case of our result by setting $\mu=0$, while by setting $M=0$ we find the cosmic string deflection angle.Given the deflection angle in the last equation, we can give a rough number estimation for the Sun case. Choosing the string linear mass density $\mu=10^{22}$ kg/m, or in natural unites $\mu=10^{-6}$, mass of the sun $M=1.48$ km with a radius $b=R=6.96 \times 10^5$ km, and finally if we take into account the fact that the sun rotates slowly, i.e $a \approx 0$, for the deflection angle we find $\hat{\alpha}=4.34 $ arc seconds. On the other hand, it is well known that the light passing near the solar limb is deflected by an angle of $1.75$ arc seconds. We conclude that the overall contribution in the deflection angle is of the order of arc seconds, in fact the deflection angle increases compered to the standard result due to the presence of the cosmic string.  

\newpage
\section{Geodesics equations}
Let us check our result for the deflection angle found in the last section by studying the geodesic equations. Furthermore we shall apply the variational principle 
\begin{equation}
\delta \int \mathcal{L} \,\mathrm{d}s=0,
\end{equation}
to the quantum improved  Kerr spacetime metric with a cosmic string, one finds the following equation for the Lagrangian 
\begin{eqnarray}\label{geo1}
\mathcal{L}&=& - \frac{\Delta_{\tilde{\omega}}}{2\Sigma}\left(\dot{t}-a \beta \sin^2 \theta \dot{\varphi} \right)^2+\frac{\Sigma}{2 \Delta_{\tilde{\omega}}}\dot{r}^2+\frac{\Sigma \dot{\theta}^2}{2}+\frac{\sin^2 \theta}{2 \Sigma}\left[a \dot{t}-(r^2(s)+a^2)\beta \dot{\varphi} \right]^2,
\end{eqnarray}
where 
\begin{eqnarray}
\Delta_{\tilde{\omega}}&=& r^2(s)+a^2-2G(r(s)) M r(s),\\
\Sigma &=& r^2(s)+a^2 \cos^2 \theta,\\
G(r)&=& \frac{G_{0}r^{3}(s)}{r^{3}(s)+\tilde{\omega} G_{0}\left[r(s)+\gamma G_{0}M\right]}.
\end{eqnarray}

As in the main text, we shall be interested in studying the deflection of planar photons by  considering $\theta =\pi/2$. Let us consider the following two constants of motion $l$ and $\gamma$, defined as \cite{Boyer}
\begin{eqnarray}\label{44}\notag
p_{\varphi}&=&\frac{\partial \mathcal{L}}{\partial \dot{\varphi}}={\frac { \left( -a\beta\,\dot{\varphi}+{\it \dot{t}} \right) a\beta}{r^2(s) } \left[ r^2(s)+{a}^{2}-{\frac {2{\it G_0}\, r^4(s) M}{ r^3(s)+\omega\,{\it 
G_0}\, \left( r \left( s \right) +\gamma\,{\it G_0}\,M \right) }}
 \right] }\\
 &-& {\frac {\beta \left[ a{\it \dot{t}}-\beta\, \dot{\varphi}\left( {a}^{2}+ r^2(s) \right)  \right]  \left[ {a}^{2}+r^2(s) \right] }{ r^2(s)}}=l,\\\notag
p_{t}&=&\frac{\partial \mathcal{L}}{\partial \dot{t}}={\frac {-a\beta\,\dot{\varphi}+{\it \dot{t}}}{ r^2(s)} \left[ {a}^{2}+ r^2(s)-{
\frac {2{\it G_0}\, r^4(s) M}{  r^3(s)+\omega\,{\it G_0}\, \left( r \left( s
 \right) +\gamma\,{\it G_0}\,M \right) }} \right] }\\
 &-& {\frac { \left[ a{\it \dot{t}}-\beta\,\dot{\varphi} \left( {a}^{2}+ r^2(s) \right) \right] a}{ r^2(s)}}=-\gamma.
\end{eqnarray}

Hereinafter it is very convenient to use a new variable $u$, which is related to the radial coordinate with the following coordinate transformation $r=1/u(\varphi)$. This gives the identity
\begin{equation}\label{iden}
\frac{\dot{r}}{\dot{\varphi}}=\frac{\mathrm{d}r}{\mathrm{d}\varphi}=-\frac{1}{u^2}\frac{\mathrm{d}u}{\mathrm{d}\varphi}.
\end{equation}

Without loss of generality one can choose $\gamma=1$ \cite{Boyer}. If we use Eqs. (\ref{geo1})-(\ref{iden}), and considering the fact that $u=u_{max}=1/r_{min}=1/b$ \cite{lorio}, this leads to $l=\beta b$. Then one finds the following differential equation, 
\begin{eqnarray}\label{33}
\frac{a^2 \beta^2}{2}-\frac{\beta^2 \Xi^2(u)}{2\zeta^2(u)}+\frac{G_0 Ma^2 \beta^2}{u^2 \Upsilon(u)}-\frac{2 G_0 Ma \beta^2 \Xi(u)}{u^2 \zeta(u) \Upsilon(u)}+\frac{G_0 M \Xi^2(u)\beta^2}{u^2 \zeta^2(u) \Upsilon(u)}+\frac{\left(\frac{\mathrm{d}u}{\mathrm{d}\varphi}\right)^2}{2 u^6 \left(u^2+\frac{1}{u^6}-\frac{2 G_0M}{u^4 \Upsilon(u)}\right)}+\frac{\beta^2}{2 u^2}=0
\end{eqnarray}
where 
\begin{eqnarray}
\Xi(u) &=& \frac{\beta \left[G_0^2 a^2 \gamma M \tilde{\omega}u^5 +G_0^2 \gamma M      \tilde{\omega} u^3+G_0 a^2 \tilde{\omega} u^4 +2 G_0 Ma^2 u^3 -2G_0 Ma u^3 b +G_0 \tilde{\omega}u^3+a^2 u^2 +1 \right]}{u^5},\\
\zeta(u) &=& \frac{\beta \left[bMu^3 G_0 \gamma \tilde{\omega} +b u^2 G_0 \tilde{\omega}+2 MG_0 (a-b)u+b\right]}{u^3},\\
\Upsilon(u) &=& G_0^2 \tilde{\omega}\gamma M+\frac{\tilde{\omega}G_0}{u}+\frac{1}{u^3}.
\end{eqnarray}

From our differential equation \eqref{33} we find
\begin{equation}
\varphi= \int_0 ^{1/b}  \frac{\mathrm{d}u}{\sqrt{2 u^6 \left(-\frac{a^2 \beta^2}{2}+\frac{\beta^2 \Xi^2(u)}{2\zeta^2(u)}-\frac{G_0 Ma^2 \beta^2}{u^2 \Upsilon(u)}+\frac{2 G_0 Ma \beta^2 \Xi(u)}{u^2 \zeta(u) \Upsilon(u)}-\frac{G_0 M \Xi^2(u)\beta^2}{u^2 \zeta^2(u) \Upsilon(u)}-\frac{\beta^2}{2 u^2}\right) \left(u^2+\frac{1}{u^6}-\frac{2 G_0M}{u^4 \Upsilon(u)}\right)}}.
\end{equation}

Note that if we consider a Taylor expansion series around $\mu$, $M$, $a$, and $\hbar$, we find the following relation 
\begin{equation}
\varphi= \int_0 ^{1/b}  A(u)  \mathrm{d}u.
\end{equation}
where
\begin{eqnarray}
A(u)=\frac{\left(4\mu+1\right)\left[ -167\,\hbar \,G_0^2 M{b}^{3}{u}^{5}+30\,{\it G_0}\,M\pi \,{b}^{
3}{u}^{3}+334\,\hbar \,{{\it G_0}}^{2}Ma{u}^{3}+30\,{b}^{3}{u}^{2}\pi -
60\,Mau{\it G_0}\,\pi -30\,b\pi
  \right]}{30\,\sqrt {-{b}^{2}{u}^{2}+1} \left( {b}^{2}{u}^{2}-1 \right) \pi }.
\end{eqnarray}

We can express the solution of our differential equation \eqref{33} in leading order terms as follows \cite{Boyer}
\begin{equation}
\Delta \varphi =\pi+\hat{\alpha},
\end{equation}
where $\hat{\alpha}$ is the deflection angle to be calculated. Furthermore the deflection angle can be calculated as follows (see for example \cite{weinberg})
\begin{equation}
\hat{\alpha}=2|\varphi(u_{max})-\varphi_{\infty}|-\pi.
\end{equation}

The deflection angle in the weak deflection limit approximation is found to be
\begin{equation}
\hat{\alpha}\simeq  4 \pi \mu +\frac{4 G_0 M}{b}+\frac{16 M G_0\mu}{b}-\frac{1336 G_0^2 M\hbar }{45 \pi b^3 }-\frac{5344 G_0^2 M \hbar \mu}{45 \pi b^3}\pm \frac{4 G_0 M a}{b^2}.
\end{equation}

This result is in complete agreement with Eq. (\ref{alp2}) found by GW method. As expected, this result clearly shows that the standard Kerr solution \cite{Boyer} is modified due to the conical topology. 
It is interesting to note that this result is not in full agreement with third order terms, such as the mixed term, $Ma\mu$, or, say, $ M a \hbar$. In particular using the geodesic approach we find the following result up to third order terms
\begin{equation}
\hat{\alpha}_{geodesics}\simeq  4 \pi \mu +\frac{4 G_0 M}{b}+\frac{16 M G_0\mu}{b}-\frac{1336 G_0^2 M\hbar }{45 \pi b^3 }-\frac{5344 G_0^2 M \hbar \mu}{45 \pi b^3}\pm \frac{4 G_0 M a}{b^2}\pm \frac{668 G_0^2 a M\hbar}{15 \pi b^4}\pm \frac{16 M a G_0\mu}{b^2},
\end{equation}
whereas the GW method gives
\begin{equation}
\hat{\alpha}_{GB}\simeq  4 \pi \mu +\frac{4 G_0 M}{b}+\frac{16 M G_0\mu}{b}-\frac{1336 G_0^2 M\hbar }{45 \pi b^3 }-\frac{5344 G_0^2 M \hbar \mu}{45 \pi b^3}\pm \frac{4 G_0 M a}{b^2}\pm \frac{1336 G_0^2 a M\hbar}{15 \pi b^4}\pm \frac{48 G_0 M a \mu}{5 b^2}.
\end{equation}

As one can see, the agreement between GB method and geodesics approach breaks down for the last two terms which can also be viewed as second order terms in mass i.e. $\mathcal{O}(M^2)$, since we assume $M\sim a$. However, as noted in Ref. \cite{kimet3}, this is not a surprising result since one needs to make a suitable choose for the vector field used in constructing the osculating Riemannian geometry. Another important point is that we have used a straight line approximation $r=b/\sin\varphi$ in calculating the deflection angle in the GBT. So another way to resolve this problem and to improve the agreement is modify the integration over the domain $\mathcal{D}_\infty$ and hopefully to find second-order correction terms.

\bigskip
\end{widetext}
\section{Conclusion}
In this paper, we have extended the GB method to the non-asymptotically flat spacetimes such as the Kerr black hole spacetime pierced by a static cosmic string. We have computed the quantum improved deflection angle of light in the weak approximation limit in the spacetime of a quantum improved Kerr black hole with a cosmic string. In the first case, we have used the GB method by introducing the quantum improved Kerr-Randers optical metric with a cosmic string and the corresponding osculating Riemannian metrics. In our second case we have used the geodesic equations and found the same result in leading order terms.  We have also shown that the deflection angle increases due to the presence of the  cosmic string. However, the main importance of GB method from the standard method used in computing light deflection is of conceptual nature. As we have shown one can find an exact result in leading order of quantum effects by integrating in a  domain \textit{outside} of the light ray which is quite remarkable result. Finally, will be interesting to see if one can find and exact result for the deflection angle, up to the third order terms, using GW method. In principle, such a thing is possible, say, by choosing an appropriate vector fields used in constructing the osculating Riemannian geometry.\\

\bigskip
\section*{Acknowledgements}
This work was supported by the Chilean FONDECYT Grant No. 3170035 (A\"{O}).

\end{document}